\documentstyle[12pt]{article}

\title{Field Theories from the Relativistic Law of Motion}

\author{Parampreet Singh\thanks{E-mail: param@iucaa.ernet.in}
\, and Naresh Dadhich\thanks{E-mail: nkd@iucaa.ernet.in}\\
{\sl Inter-University Centre for Astronomy \&
Astrophysics,} \\
{\sl Post Bag 4, Ganeshkhind, 
Pune 411 007, India .}}

\date{}
\def \n {\noindent}
\begin{document}
	
\maketitle

\begin{abstract}
 From the relativistic law of motion we attempt to deduce 
the field theories corresponding to the force law being linear and 
quadratic  in 4-velocity of the particle. The linear law leads to the vector gauge theory which could be the abelian Maxwell electrodynamics or the non-abelian Yang-Mills theory. On the other hand the quadratic law demands spacetime 
metric as its potential which is equivalent to demanding the Principle of 
Equivalence. It leads to the tensor theory of gravitational field - General Relativity. It is remarkable that a purely dynamical property of the force law leads uniquely to the corresponding field theories.
\end{abstract}
\n
 PACS: 04.20.Cv, 11.10.-z, 45.20.-d, 03.50.-z  

\n

\section{Introduction}

Until the advent of the general relativity (GR), the equations of motion 
for particle and field were two separate and independent statements, and no relation between them was sought. 
The field does however determine motion of particle by appearing on the 
right hand side of Newton's second law of motion. Whereas the equation of motion for the field is prescribed independently by the field theory without any reference to particle motion. GR brought 
forth the first instance of a relation 
between the two. In here the particle equation follows from Einstein's 
equation for gravitational field$^1$. Intuitively, it could be 
understood as follows: Gravitation is described by the curvature of 
spacetime and this fact is stated by Einstein's equation. Solution of 
this equation determines the geometry of spacetime which now incorporates
  gravitational field. Motion under gravity would then be 
free motion relative to  spacetime geometry determined by the solution 
of the field equation. This is how the particle equation is determined by 
the field equation.

 In this context, the two questions naturally arise are the following:\\
(a) Like gravitation, there exists another classical field of 
electromagnetism, why can't such a relation between the equations of 
motion of particle and field be sought for it? \\
 (b) For gravitation, the particle motion is derived from the field 
equation in GR, how about the other way round; i.e. field equation 
from particle equation.

 For the electromagnetic field, the question was first posed by 
Feynman long back and he obtained the half of the Maxwell equations 
(the source free homogeneous, the Bianchi set) by considering commutation 
relation between position and velocity. Since Feynman thought that this 
would lead to something profound and fundamental but soon realized that 
that was not the case, and  he left it at that. It has been brought 
to light in 1990 by Dyson's paper on Feynman's derivation of the Maxwell 
equations$^2$. This gave rise to some activity in terms of discussions,
rederivation and application to other 
situations$^ {3-6}$. All of them with a sole exception$^6$ concerned with the Bianchi set. Recently, we have derived the complete
set of Maxwell equations  for the non relativistic case$^7$. We begin by 
demanding the differential operator in the Newton's second law be 
self-adjoint which immediately leads to the Bianchi set. Note that in the 
familiar terms self-adjointness of the differential operator is 
equivalent to demanding the force being linear in velocity and 
derivable from a potential. Of course, it is taken that field is produced 
by certain charge, and a priori there is no reason why pseudo scalar 
charge cannot exist. Allowing for pseudo scalar charge would lead to another set of 
two (Bianchi) equations. Then the solvability of the system of equations in 
terms of the vector fields leads uniquely to the entire set of the Maxwell equations and a universal chiral relation between the scalar and pseudo scalar charges. This was all done in the non relativistic framework. In here we shall start with the relativistic equation of motion with 4-force being linear in 4-velocity and then the same arguments would readily lead to a vector gauge theory in an elegant and cogent manner. It could be the abelian Maxwell or the non-abelian Yang-Mills theory. This consideration is independent of the background spacetime which could be flat or curved. 

Note that it was the linear force law which lead to the vector gauge theory. Naturally the question arises, what field theory would the quadratic law lead to? It turns out that the quadratic law would demand spacetime metric to be its potential, which would imply the Principle of Equivalence. For force to be globally non-removable, the spacetime must be curved and the  field would then be described by the curvature of spacetime. This is how precisely gravitation is described in GR. Thus, the quadratic law would lead to GR with the proper prescription of the energy momentum tensor. 

It is remarkable that purely a dynamical consideration on the force law determines uniquely the corresponding field theories. This is in the same spirit as the Bertrand's theorem for the central force in classical mechanics which picks out only the inverse square and linear law on the demand of closed orbit. This was however restricted only to central force. In our case, the demand is on the velocity dependence of the force law, otherwise it is all general. 

\section{Maxwell's equations}

The relativistic law of motion would read as 
\begin{equation}
m \, du^i/ds = f^i 
\end{equation}  
defining the equation of motion for a test particle, where $s$ is the 
proper time. Since 4-velocity is by definition orthogonal to 4-acceleration, $f^iu_i=0$ always. 

The requirement of linearity would mean $f_a = F_{ab}u^b$. Now the orthogonality  $f_au^a=0$ would imply $F_{ab} = -F_{ba}$. Further for it to be derivable from a 
potential would require the potential to be $A_au^a$, which would lead to  
\begin{equation}
F_{ab} = \nabla_{[a} A_{b]}
\end{equation}
where $A_a = (\phi, {\bf A})$ is the gauge potential. This would satisfy the Bianchi identity,
\begin{equation}
\nabla^b \, {}^*F_{ab} = 0
\end{equation}
where ${}^*F_{ab} = \frac{1}{2} \eta_{abcd} F^{cd}$ is the dual of $F_{ab}$ and $\eta_{abcd}$ is the 4-volume form. 
Defining 
\begin{eqnarray}
{\bf B} &=& \nabla \times {\bf A} \\
{\bf E} &=& - \nabla \phi - \partial {\bf A}/ \partial t
\end{eqnarray}
then in the familiar terms, the Bianchi equation reads as
\begin{equation}
\nabla.{\bf B} = 0, ~\nabla \times {\bf E} = - \, \partial {\bf B} / \partial t
\end{equation}
where ${\bf E}$ is a polar vector and ${\bf B}$ is an axial vector.
Note that $F_{ab}({\bf E},{\bf B})$ is composed of a polar vector $\bf E$ and an axial vector ${\bf B}$. Under the duality  
transformation ${\bf F} \rightarrow\bf {}^*F$, ${\bf E}\rightarrow{\bf B}, {\bf B} \rightarrow -{\bf E}$. The source for the polar field is scalar charge while the axial field is produced by motion of charge. 

On the other hand we could as well have begun with $P_{ab}({\bf H},{\bf D})$ where the axial field ${\bf H}$ is produced by a pseudoscalar charge and polar field ${\bf D}$ by motion of this charge. That is with a field of the type ${\bf {}^*F}$ instead of ${\bf F}$. A priori there is no reason to have only $F_{ab}$ with scalar charge and not $P_{ab}$ with pseudoscalar charge. Then we will again have the Bianchi identity for $P_{ab}$,
\begin{equation}
\nabla^b\, {}^*P_{ab} = 0 
\end{equation} 
leading to
\begin{equation}
\nabla.{\bf D} = 0, ~\nabla \times {\bf H} =  \, \partial {\bf D} / \partial t.
\end{equation}

Note that $P_{ab}$ has the same structure as that of  
${}^*F_{ab}$. We now have four equations in four vector fields which cannot 
be solved because to specify a vector field both its divergence and curl must be given. That is for each vector, there should be two equations. We have only four equations for four vectors while there should be twice as many. The only way this system could be solved is to reduce the vectors from four to two by prescribing a linear relation between the two sets. That is assuming a linear relation between ${}^*F_{ab}$ and $P_{ab}$ because they are of the similar type. Thus we write
\begin{equation}
P_{ab} = \alpha \,{}^*F_{ab}
\end{equation}     
where $\alpha$ is a proportionality constant. Then eq.(7) would become
\begin{equation}
\nabla^bF_{ab} = 0
\end{equation}
because $** = -1$. If we identify the polar field ({\bf E}) and the axial
field ({\bf B}) as the electric and magnetic fields produced by some
charge distribution (not in our consideration) then this is the other set of the Maxwell equations. We 
have thus deduced the complete set of the field equations for vacuum.  Of course the above 
linear relation would also imply that the two kinds of charges are 
no longer independent and they must also be related by a chiral relation of the type,
\begin{equation}
q_p = q_s \tan\theta
\end{equation}
where $\theta$ is 
a universal constant for a given family of particles. Thus there could exist only one kind of 
charge, calling it electric or magnetic is simply a matter of giving 
name$^7$. In a different context, the above relation was also considered by Schwinger$^8$. 

Thus the linear relativistic law of motion leads uniquely to the classical electrodynamics.  The above deduction would run through similarly with the appropriate derivative operator for the non-abelian gauge potential as well leading to the Yang-Mills equations. The linear law thus leads to the vector gauge theories. Further, in our consideration the background metric did not enter specifically  and hence it would remain valid for both flat as well as curved spacetime. 

\section{Einstein's equations}

For the quadratic law, we must have
\begin{equation}
f^a = -T^a_{bc} \, u^b \, u^c. 
\end{equation}
For this equation to be derivable from an Action principle, the Lagrangian must have a quadratic 
term in velocity which should be a scalar and it could be written as
\begin{equation}
L = \frac{1}{2} \hspace{0.15cm}  p_{ab}u^au^b .
\end{equation}
 This Lagrangian on variation would give the equation of motion,
\begin{equation}
 p_{ab}\frac{du^b}{ds} = -P_{bc,a}u^bu^c,
\end{equation}
 where 
\begin{equation}
P_{bc,a} = (1/2)(p_{ba,c} + p_{ac,b} - p_{bc,a}).
\end{equation}

 Here something remarkable has happened. That the equation of motion for the quadratic law is independent of mass of the particle. This is the property which is known as the Principle of Equivalence (PE). It is important to note that PE is the characterizing feature of the quadratic law and not necessarily of gravity alone. Thus all forces that are quadratic in velocity would obey PE. 

 The 4-force should be orthogonal to the 4-velocity, which would imply,
\begin{equation}
u^a p_{ab}\frac{du^b}{ds} = -\frac{1}{2} \, p_{ba,c}u^bu^cu^a = 0
\end{equation}
because of symmetry in all the three indices. This could be true only if $p_{ab}$ is antisymmetric, which is impossible. The only way the equation can be made to have some sensible meaning is that $p_{ab} = g_{ab}$ define the spacetime metric. Then the equation would become the geodesic equation, $\dot u^a = u^b \nabla _b u^a = 0$ for the metric $g_{ab}$ with $P_{ab,c}$ defining the Christoffel symbols. Then the 4-acceleration would be by definition orthogonal to 4-velocity.
 
 We have thus reached the fundamental and profound conclusion that the
quadratic force requires the spacetime metric as its potential. It is
profound for the reason that it is the first instance of a force law
making a demand on the spacetime geometry which marks an important break from the classical paradigm of given inert spacetime background. So far we have made no reference to any particular field. A number of inferences readily follow:
(a)It now ceases to be an external force and instead becomes a property of the spacetime geometry, which would be felt by timelike as well as null particles alike.(b) It must therefore link to all particles and energy forms
alike including photons. That is, it has the universal linkage. (c)Since the force would now arise through the Christoffel symbols, which
would be globally non-zero only when spacetime is curved having non-zero
Riemann curvature. If force is to be globally non-removable, then   spacetime must be a curved Riemannian manifold. (d) Another way to state PE is that the equation of motion must be free of mass of particle. When that happens, force could only be derived from spacetime geometry. 

In the relativistic law of motion, null particles must also be included. The characteristic feature of null particles is that they only respond to the spacetime geometry and not to any external force, and their propagation vector is null $u^au_a = 0$. Any field which is to be shared by both ordinary and null particles has to be the property of spacetime geometry. We turn the question around to ask when could equation of motion be independent of mass of particle? The answer would be only when motion is purely driven by the spacetime metric. 

All this is true for any force law which is quadratic. This is 
another matter that gravitation and inertial forces are the only known
examples of such a force law. The former is globally non-removable while
the latter is by a suitable coordinate transformation. This is why inertial forces were called fictitious because they could be removed by a suitable choice of spacetime geometry. What removal here means is incorporation into geometry of spacetime, which would now act as their potential. The distinguishing criterion between inertial forces and gravitational field is the Riemann curvature of spacetime. The metric that incorporates inertial forces would have vanishing curvature while that for gravitation would have non-zero curvature. 

 Since the spacetime is a Riemannian manifold, its  curvature would satisfy the
differential Bianchi identity
\begin{equation}
R_{a(bcd;e)} = 0 
\end{equation}
which on contraction would as is well-known
give 
\begin{equation}
\nabla_bG^{ab} = 0
\end{equation}
where $G^{ab}$ is the Einstein tensor. It integrates to give
\begin{equation}
G^{ab} = \kappa \, T^{ab} + \Lambda \, g^{ab}
\end{equation}
with $T^{ab}$ being divergence free. Now if we identify $T^{ab}$ with the
energy momentum tensor of matter/energy distribution, $ \Lambda $ with the cosmological constant and
$\kappa$ involving gravitational constant, then this is the familiar
Einstein equation with the proper Newtonian limit. Note that the two terms on the right are the
sources for the curvature of spacetime which incorporates in its geometry the quadratic force law of the gravitational field. We have thus deduced the Einstein field equation for gravitation with sources. The vacuum is defined by $T_{ab} = 0 = \Lambda$. 

The dynamical characterization of electromagnetic and gravitational
fields is thus by the linear and quadratic character of the force law. In this
sense gravity is the generalization of the classical electrodynamics or
rather the Yang-Mills field. From the field theoretic viewpoint it is
the zero mass spin 1 photon  characterizes the electromagnetic
field. This is essentially brought about by studying
the effects of rotation on the transverse components of the vector 
potential$^9$. The existence of a 4-vector potential for the field
determines that the mediating particle must be of spin
1 and since the interaction is long range, it should be massless. In our consideration, it is the linear character of the force law that
demands a 4-vector potential. This is how both classical and quantum  
characterizations meet at the level of gauge potential. 
 
 In the case of gravitation, the quadratic law requires the spacetime to
be curved and the metric itself plays the role of potential. We can construct
an action for fields and by taking the weak field limit and
write the plane wave solutions as  $h_{ab} = \epsilon_{ab} \,  exp(i k.x)$,
where $\epsilon_{ab}$ is the polarization tensor$^{10}$.
Then following the similar procedure
it can be shown that only the transverse components of  $\epsilon_{ab}$ 
are physically relevant which would correspond to the spin 2 massless graviton.

In essence our path is reverse from the one followed in field theories$^{11}$.
There in order to  have a field theory of massless particles with spin, we write the field in terms of creation and annihilation operators. Then the demand of the Lorentz and gauge invariance restricts the 
 force field for spin 1 field  to an antisymmetric tensor $F_{ab} = \partial _a  A_b - {\partial}_b A_a$ with a coupling current $J ^a$ satisfying the conservation equation 
$\partial _a J^a = 0$. This then leads to the Maxwell's equations in vacuum.
Similar demand for the field of a spin 2 massless particle would lead to a 
tensor $R_{abcd}$ having the same symmetry properties of the Riemann tensor.
To identify this field with that of in the linearized gravity we require the
introduction of a symmetric tensor $h_{ab}$ and a coupling current $T^{ab}$
satisfying the conservation equation $\partial _a T^{ab} = 0$. Note that although the above construction was in the linear regime, one can recover full GR by lifting the restriction of linearity. The important point is that no sooner 
we strike the Riemann curvature tensor, we have landed in the Einstein's 
theory of gravitation.

\section{Discussion}

 Let us now reflect on the route  from field to particle motion. Given the field equation, how do we deduce the particle equation ? For the vector gauge theory, we are given a 4-vector potential. Its interaction with the particle should be given by a scalar which could only be formed by the scalar product of the 4-potential with the 4-velocity of particle (which is the only 4-vector particle could have of its own). This would lead to an interaction term of the form
$(A_i u^i)^n$ and for a gauge invariant theory $n$ can not be different from 1. This would then  lead to the familiar particle equation of motion. Alternatively we could have demanded that the field which is a 2-form and appears linearly in the particle equation. In the case of gravity, the field equation yields a curved spacetime with its metric serving as potential. The particle equation would then follow from the invariant constructed of the metric and 4-particle velocity. Since it is quadratic in velocity, the equation would be free of particle's mass and would automatically obey PE. It would be given by the geodesic equation. Though the Newtonian equation of motion for particle is also independent of mass but it does not obey in the strict sense PE because the equation does not refer to null particles. For the relativistic equation, the requirement of mass independence is equivalent to PE as well as to spacetime metric serving as potential for motion. 
   
 In summary, we have illustrated a simple and elegant method to deduce the
corresponding field theories to linear and quadratic force law. The linear law yields the vector gauge theories; Maxwell theory for the abelian and the Yang-Mills theory for the non-abelian case. This deduction is independent of the spacetime background which could be flat or curved. The quadratic law marks a fundamental break by requiring spacetime metric to be its potential and it leads to the tensor theory. This is the most important statement. Since gravitation also shares this property, the Einstein equation for gravitation could be deduced. It is always insightful and illuminating to see inter relations between various physical concepts and statements. This exercise is an attempt to understand the relation between motion of field and particle. \\

\noindent
Acknowledgment: Over a period we have benefited by discussions with many colleagues, particularly Juergen Ehlers and Ghanshyam Date. We thank them warmly.
PS thanks Council for Scientific \& Industrial Research for grant 
number: 2 - 34/98(ii)E.U-II.

\end{document}